# Study of ($^3$He, *t*) charge exchange reactions to isobaric analog states in inverse kinematics

Zhixuan He[1,a], Wenjuan Bu[1], Chaoyuan Xiao[3], Meng Li[4], Herun Yang[2], Bitao Hu[1], Yi Zhang[1,b]

[1] *School of Nuclear Science and Technology, Lanzhou University, 222 South Tianshui Road, Lanzhou, 730000, Gansu Province, China*

[2] *Institute of Modern Physics, Chinese Academy of Sciences, 509 Nanchang Road, Lanzhou 730000, Gansu Province, China*

[3] *China Nuclear Power Technology Research Institute Co., Ltd., 1001 Shangbu Middle Road, Shenzhen 518000, Guangdong Province, China*

[4] *Department of Oncology, The Second Xiangya Hospital, Central South University, No.139 Renmin Road Central, Changsha 410011, Hunan Province, China*

**Abstract**: The transition between isobaric analog states (IAS) in the ($^3$He, *t*) charge exchange reaction presents a unique opportunity to access the isospin structure of the nuclei. In this study not only the Fermi transition but also the Gamow-Teller (G-T) transition of the IAS reaction were investigated for the $^{13,14}$C($^3$He, *t*) and $^{17,18,19,20}$O($^3$He, *t*) reactions, in order to explore the neutron number dependence of the IAS reaction for the light neutron-rich nuclei. It was found that the G-T type IAS reaction also exhibited a significant dependence of the transition strength on the neutron number and the angular momentum configuration of the nuclei. Additionally, the inverse kinematics was also discussed for extracting the yields of the interested reaction channels in the proposed experiments on radioactive beams. The calculated triton yields demonstrated the capability of the proposed experiments to obtain meaningful results.

## 1 Introduction

Direct nuclear reactions at intermediate energies offer a clean and convenient way to observe dynamical properties in nuclei. Through the charge exchange (CE) reaction at this energy region, we can access not only the isospin structure of the nucleus [1-3], but also the isovector interaction between the projectile and target [4]. Within the distorted wave Born approximation (DWBA), the 'double-folding' analysis has shown that the isoscalar and isovector terms of the Lane-form potential represent the rescaled isoscalar and isovector densities of the target nuclei, respectively [5]. Furthermore, not only the interaction strength but also the shape of the isovector potential plays a key role in the CE reaction [6]. Extensive work had been done to reliably understand the Lane-form optical potential with a microscopic framework [7-9]. As the isovector term of the Lane-form potential shows how different the protons and neutrons behave during a reaction, and the isovector density shows how different the protons and neutrons distribute inside a nucleus, it turns out that both originate from the same mechanics termed charge symmetry [10-11]. Due to the neutron-proton asymmetry of the target nucleus, $(N-Z)/A$, the isovector potential is normally small compared with the isoscalar counterpart. Even though extensive data on CE reactions had been accumulated, it was still not enough for an optical potential to give a satisfactory description of the experiment [12]. Rare-isotope beam nuclear reactions in inverse kinematics would have significant advantages, especially in the light-nucleus region.

[a] E-mail: hezhx21@lzu.edu.cn
[b] Corresponding author, e-mail: yizhang@lzu.edu.cn

Taking the spin dimension into account, the CE reactions can be grouped into two types: spin-flip (Gamow-Teller or G-T) transition and non-spin-flip (Fermi or F) transition, respectively. In the G-T transition, the spin of target nuclei changes by 1 ($\Delta S = 1$), and the cross section contains the off-diagonal elements of the transition matrix. Therefore, the G-T transition has been studied in experiments extensively to access the spin-dipole (SD) nuclear matrix element, which is useful to access the isovector-spin resonance of the nuclei [13-16], and is of special interest in searching the neutrinoless double $\beta$ decay [17-18]. On the other hand, the Fermi transition where $\Delta S = 0$ is a pure isospin excitation. In addition, there is a special excitation ideal to access the isovector structure, which is the excitation between the isobaric analog states (IAS). The nuclei in the initial and final states of the IAS reaction have a similar structure, where only one nucleon is replaced. Since the total angular momentum and the parity of the initial and final states are the same in the IAS reaction, the transition can be treated exactly as an elastic scattering except for the isospin orientation of the flipped nuclei. Thus, it could be employed as a primary tool to study the isospin symmetry of the nuclei in the initial and final states, such as the nuclear symmetry energy and the neutron skin [19]. In both theory and experiment, there were several recent works to analyze the isovector densities and diffuseness of the target nucleus based on the isovector term of the optical potential and the double-folding formalism [20-22]. The experimental data of the ($^3$He, $t$) scattering to the IAS of $^{90}$Zr and $^{208}$Pb at $E_{lab}$ = 420 MeV have been studied to deduce neutron skin values for these two nuclei [20]. A neutron skin $\Delta R_{np} \approx 0.16 \pm 0.04$ fm was obtained for $^{208}$Pb and $\Delta R_{np} \approx 0.09 \pm 0.03$ fm for $^{90}$Zr. In addition, SD excitations of $^{90}$Zr are studied by the $^{90}$Zr($p, n$) and $^{90}$Zr($n, p$) CE reactions at 300 MeV, and the neutron skin thickness of $^{90}$Zr is determined to be $\Delta R_{np} \approx 0.07 \pm 0.04$ fm [23]. These successful attempts set the stage for the extraction of neutron radii and neutron skin thicknesses using CE reactions.

Compared with the basic ($p, n$) process, the ($^3$He, $t$) process is more sensitive to the outer, rather than the inner, structure of the target nuclei [24-25] and was measured with a significantly higher resolution [26-27]. A series of theoretical and experimental studies on ($^3$He, $t$) reactions had been carried out for several nuclei, relying on the combination of $^3$He beam and stable or long-lived isotopes as the targets [28-30]. As a complementary, with a radioactive beam, the CE reaction in inverse kinematics would offer a great opportunity to study the isospin structure of the nucleus far from the $\beta$-stable line [31-32]. It would be especially interesting to observe the IAS reactions of the neutron-rich light nuclei in the ($^3$He, $t$) process, as the process would favor the isovector structure of the nucleus surface and isovector potential would be stronger due to the factor of $(N-Z)/A$. Therefore, a ($^3$He, $t$) experiment plan was proposed on the radioactive beams of the Heavy Ion Research Facility in Lanzhou (HIRFL) [33-34]. This would be an excellent opportunity to investigate the structure of unstable neutron-rich nuclei, especially for neutron drip line nuclei, where the proton-neutron asymmetry is significant.

Besides, for the light mirror nucleus such as $^{7}$Li-$^{7}$Be, $^{15}$N-$^{15}$O, and $^{17}$O-$^{17}$F, the possible cluster effect is another charming topic based on a similar experiment configuration [35-38].

In observations of IAS reactions, it is necessary to carefully choose the kinematics of the reaction to precisely identify the final state of the outgoing nuclei. As in the center-of-mass frame, the IAS reaction mainly distributes in the small-angle region, observing them in inverse kinematics could take advantage that those small-angle scattering events in the center-of-mass frame, who correspond to the large-angle scattering events in the laboratory frame, are far from the beam and clean to observe. Furthermore, in inverse kinematics, the energy of the recoiled nuclei varies rapidly with its scattering angle. This strong dependence can be employed as a powerful selection rule and calibration tool to make precise measurements.

Even through, we noted that in previous works the observation was limited to the even-even nucleus [20-22,39-40]. In these cases, only the strength of the Fermi transition in the IAS reaction was proved to have an unambiguous correlation with the isovector structure. Meanwhile, in the odd-A case the IAS reaction might be contributed by both the Fermi and G-T transitions, which is more complex. It is necessary to investigate both transitions in the odd-A case, so as to separate their contributions in the same reaction. Moreover, for the radioactive beam experiment, there are several issues that need to be investigated in detail to demonstrate the feasibility of the experiment. In this work, the cross sections and their angular distributions for several CE reactions were calculated theoretically. The kinematic variables, including the kinetic energy and angle, have also been optimized for the target and beam conditions in inverse kinematics [33-34,41]. The target absorption of the recoiled triton can be effectively overcome by choosing the beam energy appropriately. The Fermi transition and G-T transition were extracted from other contributions in the IAS reaction by the demonstrated multipole decomposition analysis (MDA) [42-43], according to respective cross section and angular distribution.

In Section 2, the theoretical framework is described. In Section 3, the cross sections of the CE reactions are calculated, the correlation between the cross sections and the nuclear radii is investigated, and the experimental plan is discussed. In Section 4, the calculation is summarized and the conclusion is presented.

## 2  Theoretical framework

In this work the Double Folding Model (DFM) in the DWBA [44-46] was employed to analyze the CE reaction. In this framework, incoming and outgoing particles are regarded as the plane wave and spherical wave distorted by the mean field of the target nucleus, respectively. The transition matrix can be expressed as:

$$T_{fi} = \langle \chi_f | F(s) | \chi_i \rangle, \qquad (1)$$

where $\chi_{i(f)}$ is the distorted wave of the initial (final) state and $F(s)$ is the form factor. Then the differential reaction cross section can be written as [47]:

$$\frac{d\sigma}{d\Omega} = \left(\frac{\mu}{2\pi\hbar^2}\right)^2 \frac{k_f}{k_i} |T_{fi}|^2, \qquad (2)$$

where $\mu$ is the reduced mass, while $k_{i(f)}$ is the incoming (outgoing) wave number. The form factor $F(s)$ describes the interaction between the projectile and target comprehensively. According to the DFM, the transition densities of the projectile-ejectile and target-residue systems are 'folded' (integrated) with the effective nucleon-nucleon (*NN*) interaction to produce the form factor. The 'folding' process is expressed as an integral [48]:

$$F(s) = \int \rho_{ab}(r_p) V_{eff}(s, r_p, r_t) \rho_{AB}(r_t) \, dr_p dr_t, \quad (3)$$

where $\rho_{ab}$ is the transition density for the target-residue system, $\rho_{AB}$ is the transition density for the projectile-ejectile system, and $V_{eff}$ is the nucleon effective interaction between the projectile and target. The transition density is defined as:

$$\rho_{LSJ} = \Sigma_{np} \langle f ||a^\dagger a|| i \rangle [\phi^* \phi], \quad (4)$$

where $i(f)$ is the initial (final) state, $a$ and $a^\dagger$ are the creation and annihilation operators, respectively, and $\phi$ is the single-particle radial wave function. $\langle f ||a^\dagger a|| i \rangle$ is defined as the one-body transition density (OBTD).

It is worth noting that the form factor depends on the type of transition. Different transitions are identified by the angular momentum coupling of the projectile and target. With the total angular momentum of the projectile (target) being defined as $J_{p(t)}$, the total angular momentum transfer in the relative coordinate can be expressed as $J_r = J_p + J_t$. In case of the spin-orbit term is zero and only are the central and tensor forces considered, $J_r$ is equal to the orbital angular momentum transfer $L$ [29,49]. The $J_r J_p J_t$ can be written as the *LSJ* transfer language, where $J_r = L$, $J_p = S$, and $J_t = J$ [49]. The transition with $L = 0$ and $S = 1$ is the G-T transition (in *LSJ* form as 011), while the transition with $L = 0$ and $S = 0$ is the Fermi transition (in *LSJ* form as 000). By selecting and adjusting *LSJ* combinations, form factors for different reaction channels and transitions can be calculated.

In this work, the cross sections of ($^3$He, $t$) IAS reactions for C and O isotopes are calculated with the FOLD package [50]. The FOLD package includes three parts: WSAW, FOLD, and DWHI.

In the FOLD code, the form factor is calculated by employing Eq. **(3)** and Eq. **(4)**. The OBTDs, single-particle radial wave functions, and effective interaction should be entered into it. In the FOLD code the OBTD is included in the '*Z*-coefficient' convention [51]:

$$Z_{\Delta J, \Delta T} = a_{\Delta J, \Delta T} \langle T_i T_{zi} \Delta T \Delta T_z | T_f T_{zf} \rangle \sqrt{\frac{(2\Delta T+1)}{(2J_i+1)(2T_f+1)}}, \quad (5)$$

where $Z_{\Delta J, \Delta T}$ is the Z-coefficient, $a_{\Delta J, \Delta T}$ is the OBTD calculated by the shell-model code NuShellX [52], and $\langle T_i T_{zi} \Delta T \Delta T_z | T_f T_{zf} \rangle$ is the Clebsch-Gordan (C-G) coefficient [53]. In the NuShellX calculation, the CKPOT [54] and USDA [55] effective interactions are employed for the *p*-shell space cases and the *sd*-shell space cases, respectively. It is easy to calculate $Z_{\Delta J, \Delta T}$ from $a_{\Delta J, \Delta T}$, with the total angular momentum ($J$) and isospin ($T$ and $T_z$) defined. The single-particle radial wave functions $\phi$ in Eq. **(4)** are obtained by the WSAW code with Woods-Saxon form conveniently. In particular, for $^3$He and $t$, the single-particle wave functions are from quantum Monte Carlo simulations [29,56] rather than WSAW. On the other hand, the Love-Franey *NN* effective interaction [57-58], based on phenomenological *NN* scattering amplitudes at several bombarding energies between 50 and 1000 MeV/nucleon, is employed as $V_{eff}$ in Eq. **(3)**. To apply this *NN* interaction for a nucleon-nucleus (*NA*) system, a transformed is employed [57] as shown in Eq. **(6)**:

$$t_{NA} = \frac{\epsilon_0^2}{\epsilon_p \epsilon_t} t_{NN}, \quad (6)$$

where $\epsilon_{p(t)}$ is the total energy of the projectile (target) nucleon in the *NA* system, $\epsilon_0$ is the total energy of the incident nucleon in the *NN* system, $t_{NN}$ is the *t*-matrix of the *NN* scattering, and $t_{NA}$ is the *t*-matrix of the *NA* scattering.

The differential cross section is obtained from a DWBA calculation as shown in Eq. **(1)**,

by employing the `DWHI` code. The form factors in Eq. **(1)** are calculated with the `FOLD` code, while the distorted wave functions are generated with the `WSAW` code. The optical potential parameters can be determined by a similar collision (see Table 1). For C isotopes, the potential parameters are obtained from the fit of measured differential cross sections of $^3$He and $^{12}$C elastic scattering data at 443 MeV [59]. For O isotopes, the parameters are from the fit of measured differential cross sections of $^3$He and $^{16}$O elastic scattering data at 420 MeV [26]. The parameters of the outgoing channel are set equal to those of the incoming channel, except for the real and imaginary depths. The potential depths are scaled by a common factor of 0.85, which is assuming that the potential depth of the outgoing channel is 85% of the incoming channel one [46]. Combining the potential parameters, form factors, spin and isospin information, the differential cross sections at a given scattering angle and kinematic energy are calculated by summing the contributions of various transitions of interest in CE reactions.

**Table 1** Woods-Saxon optical potential parameters provided by the elastic scattering experiments

|  | $V_R$ / MeV | $r_R$ / fm | $a_R$ / fm | $W_I$ / MeV | $r_I$ / fm | $a_I$ / fm |
|---|---|---|---|---|---|---|
| $^{12}$C + $^3$He | 19.73 | 1.592 | 0.705 | 37.76 | 0.989 | 0.868 |
| $^{12}$N + $t$ | 16.80 | 1.592 | 0.705 | 32.10 | 0.989 | 0.868 |
| $^{16}$O + $^3$He | 22.08 | 1.540 | 0.740 | 42.66 | 0.890 | 0.960 |
| $^{16}$F + $t$ | 18.81 | 1.540 | 0.740 | 36.16 | 0.890 | 0.960 |

## 3 Results and discussion

3.1 Calculation of IAS reaction cross section

In this work, several ($^3$He, $t$) reactions involving C and O isotopes are analyzed, including $^{13}$C($^3$He, $t$)$^{13}$N, $^{14}$C($^3$He, $t$)$^{14}$N, $^{17}$O($^3$He, $t$)$^{17}$F, $^{18}$O($^3$He, $t$)$^{18}$F, $^{19}$O($^3$He, $t$)$^{19}$F, and $^{20}$O($^3$He, $t$)$^{20}$F.

In order to verify the accuracy of our calculation, the differential cross sections of $^{13}$C($^3$He, $t$)$^{13}$N and $^{18}$O($^3$He, $t$)$^{18}$F were compared with experimental results. For $^{13}$C($^3$He, $t$)$^{13}$N reaction, both the ground state and an excited state (3.578 MeV) of the produced $^{13}$N nuclei are taken into consider. The angular distribution of the differential cross sections of the $^{13}$C(1/2$^-$, g.s.) → $^{13}$N(1/2$^-$, g.s.) reaction channel **(a)** and $^{13}$C(1/2$^-$, g.s.) → $^{13}$N(3/2$^-$, 3.578 MeV) reaction channel **(b)** is shown in Fig. 1. For each channel, the contributions from different angular momentum combinations are shown as the dashed lines, while the combined differential cross sections are shown as the solid lines. The triangular data points are experiment measurements, in which the $^3$He beam with an incidental energy of 150 MeV/nucleon was employed [2]. For compatibility, the kinematic conditions are the same in the center-of-mass frame in the calculation. Within the range of scattering angle below 30.0°, the `FOLD` calculation and the experimental values are almost identical in shape, while the calculated values are slightly larger than the experimental values. It is similar to the $^{18}$O($^3$He, $t$)$^{18}$F(0$^+$, 1.592 MeV) case shown in Fig. 2, where the beam energy is 140 MeV/nucleon. It is clear that the shape of the experimental angular distribution [60] is also in agreement with the calculated values in the range of 0° ~ 7.5°. The consistency of differential cross sections demonstrates the reliability of the calculations. It is worth noting that in the $^{18}$O($^3$He, $t$)$^{18}$F case the cross section is only contributed by the IAS reaction mediated by the Fermi excitation, whose strength has a definite correlation with the neutron radii of the target nuclei, as well established in literature [20-22]. On the contrary, in the $^{13}$C($^3$He, $t$)$^{13}$N case both Fermi type (the 000 term) and G-T type (the 011 term) IAS reactions contribute to the cross section. These two types of excitation processes can be extracted separately according to their different azimuthal distribution, as shown in Fig. 1(a). It will be interesting to verify whether the IAS reaction mediated by the Fermi transition in odd-A isotopes also has correlations, as in the even-even case, and whether the IAS reaction mediated by the G-T transition has similar correlations with the isovector structure of the odd-A nucleus. In this work, a model-based investigation should shed light on these topics.

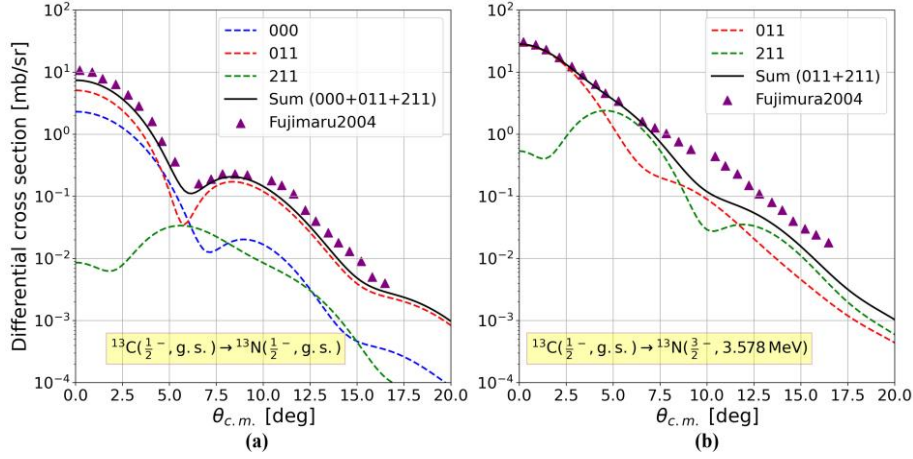

**Fig. 1** Angular distributions of the CE reaction cross sections with the 150 MeV/nucleon $^3$He beam bombarding the $^{13}$C target, including the $^{13}$C(1/2$^-$, g.s.) → $^{13}$N(1/2$^-$, g.s.) reaction channel **(a)** and $^{13}$C(1/2$^-$, g.s.) → $^{13}$N(3/2$^-$, 3.578 MeV) reaction channel **(b)**. The calculated curves are multiplied by a scale factor (2.5) to fit the experimental data

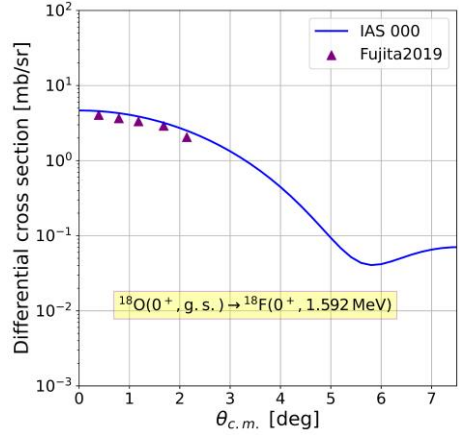

**Fig. 2** Angular distribution of the differential cross section of the $^{18}$O(0$^+$, g.s.) → $^{18}$F(0$^+$, 1.592 MeV) reaction channel with the 140 MeV/nucleon $^3$He beam bombarding the $^{18}$O target. The same scale factor (2.5) is multiplied for the calculated curve to fit the experimental data

Besides the two reactions mentioned above, in this work several other CE IAS reactions are studied, including $^{14}$C($^3$He, $t$)$^{14}$N, $^{17}$O($^3$He, $t$)$^{17}$F, $^{19}$O($^3$He, $t$)$^{19}$F, and $^{20}$O($^3$He, $t$)$^{20}$F. All reactions are divided into two groups according to the angular momentum of the initial and final states. The first group includes reactions of 0 → 0 transition, such as $^{14}$C → $^{14}$N, $^{18}$O → $^{18}$F, and $^{20}$O → $^{20}$F, which only are mediated by the Fermi transition, shown in Fig. 3**(b)**, **(c)** and **(d)**. The second group includes reactions with $J_i = J_f \neq 0$, such as $^{13}$C(1/2$^-$) → $^{13}$N(1/2$^-$), $^{17}$O(5/2$^+$) → $^{17}$F(5/2$^+$), and $^{19}$O(5/2$^+$) → $^{19}$F(5/2$^+$), containing Fermi, G-T and other transitions, shown as Fig. 3**(a)**, **(d)** and **(e)**.

The initial state is assumed to be the ground state. The final state may be the ground or excited state. As shown in Fig. 3**(a)**, $^{13}$C and $^{13}$N are mirror nuclei with $|T_{zi}| = |T_{zf}|$. The ground states of the two nuclei are two members of the isobaric multiplet, as are the $^{17}$O and $^{17}$F. On the other hand, for reactions with $^{14}$C and $^{18,19,20}$O isotopes, the isospin of the ground state of the ejectile nuclei is $T_{f0} = T_i - 1$. So only the excited state of the ejectile nuclei with proper isospin could be identified as the IAS, as shown in Fig. 3**(b)**, **(d)**, **(e)** and **(f)**. The lowest excitation energy of IAS reactions for $^{14}$N is 2.69 MeV, and for $^{18,19,20}$F are 1.592 MeV, 7.904 MeV, and 6.979 MeV, respectively, according to a `NuShellX` calculation. Cross sections of IAS reactions with higher excitation energy levels, such as 8.782 MeV for $^{13}$N, 16.364 MeV for $^{14}$N, and 5.865 MeV for $^{18}$F, are orders of magnitudes smaller and are neglected for brevity.

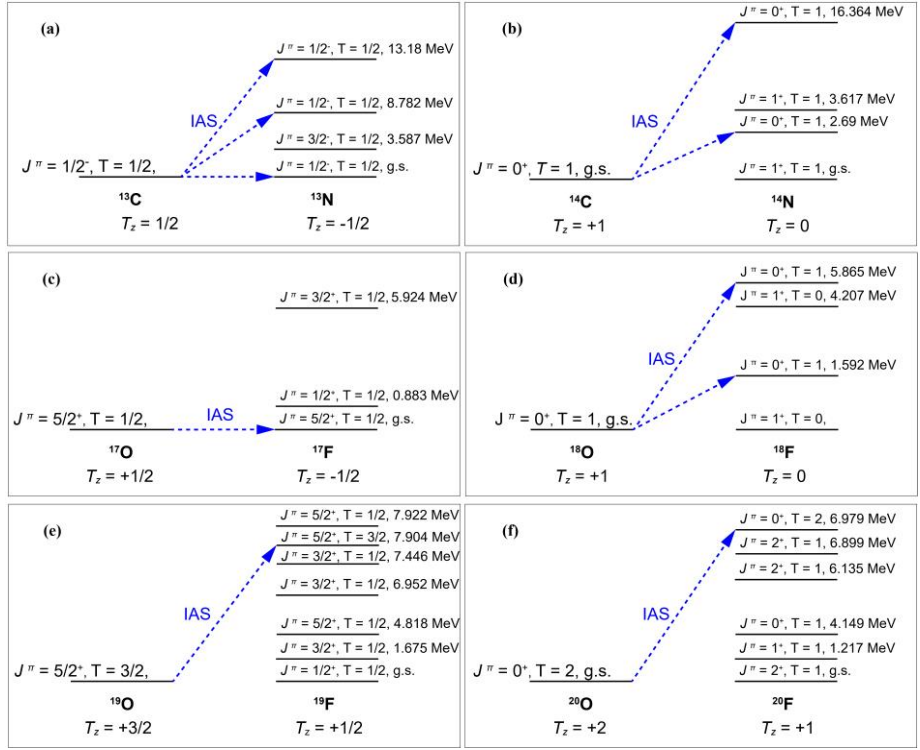

**Fig. 3** Transitions between initial and final states in the CE reaction of $^{13}$C → $^{13}$N **(a)**, $^{14}$C → $^{14}$N **(b)**, $^{17}$O → $^{17}$F **(c)**, $^{18}$O → $^{18}$F **(d)**, $^{19}$O → $^{19}$F **(e)** and $^{20}$O → $^{20}$F **(f)**. Some energy levels are omitted, and blue arrows refer to IAS reactions

Differential cross sections for $^{13,14}$C($^3$He, $t$) and $^{17,18,19,20}$O($^3$He, $t$) IAS reactions at 515 MeV/nucleon, including various transition types and their sum, are shown in Fig. 4 and 5. Due to the mixing of the G-T transition in the cross section in the odd-$A$ case, there is no apparent correlation between the mass number $A$ of the target nuclei and the total cross section of the IAS reaction for a particular isotope chain. However, in the case where only the Fermi transition is counted in the cross section, the differential cross sections show similar azimuthal dependence and an apparent positive correlation with the mass number $A$, both in the even-$A$ and odd-$A$ isotopes of a specific isotope chain, as shown in Fig. 6. In addition, the IAS cross section for counting only the G-T transition shows the same correlation, as shown in Fig. 7.

This can be attributed to the fact that both excitation types are correlated with the nuclear structure by the same mechanism but with different angular momentum couplings. Thus, it is predictable that a precise and unified description of the IAS reactions for both even-$A$ and odd-$A$ nuclei will be useful for extracting the isovector structure of exotic nuclei from CE reactions.

This is even more interesting when considering the relative variation between the Fermi and G-T transitions in different odd-$A$ isotopes of an element. As shown in Fig. 5, in the $^{17}$O case (a), the two types of transitions give rough the same contributions to the differential cross section, while in the $^{19}$O case (b) the Fermi transition apparently contributes more than the G-T transition. In other words, in the ($^3$He, $t$) process, the G-T excitation is more sensitive to the angular momentum configuration of the involved target nuclei on ($2p_{1/2}$ for $^{15}$O versus $3d_{5/2}$ for $^{17}$O). This feature would be valuable when probing the probabilistic clustering effect in the IAS reactions, since the clustering effect can be intuitively treated as a 'normal' energy state superimposed by a clustering state. Another notable advantage of employing G-T transition in ($^3$He, $t$) process rather than in ($p$, $n$) process is that ($^3$He, $t$) process is more sensitive to the surface structure of the target nuclei, where the nuclear clustering takes place. In this work the differential cross sections for $^7$Li($^3$He, $t$)$^7$Be, $^{15}$N($^3$He, $t$)$^{15}$O and $^{17}$O($^3$He, $t$)$^{17}$F IAS reactions at 515 MeV/nucleon are calculated under the theoretical framework mentioned above, as shown in Fig. 8. As the framework does not account for the clustering effect of the target nucleus, the results in Fig. 8 merely demonstrate a possible way to extract the strength of the G-T transition from the IAS cross section by partial wave analysis. The potential clustering effect can be observed by comparing the relative variation of the G-T type and Fermi type transitions within each isotope chains for $^7$Li, $^{15}$N, and $^{17}$O, respectively.

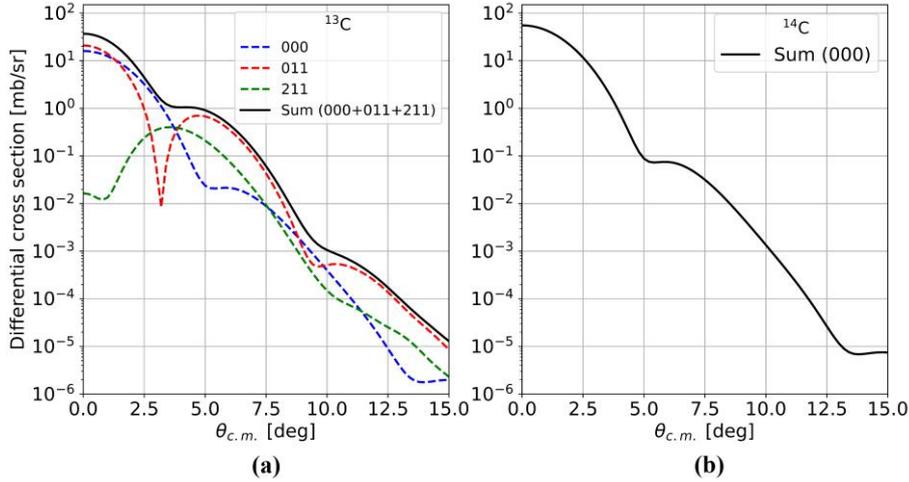

**Fig. 4** Differential cross sections for $^{13}$C($^3$He, $t$)$^{13}$N (**a**) and $^{14}$C($^3$He, $t$)$^{14}$N (**b**) IAS reactions at 515 MeV/nucleon versus scattering angles in the center-of-mass frame. The blue dashed line represents the Fermi transition, the green dashed line represents the G-T transition, and the black solid line represents the sum in (**a**). The black solid line represents the pure Fermi transition in (**b**)

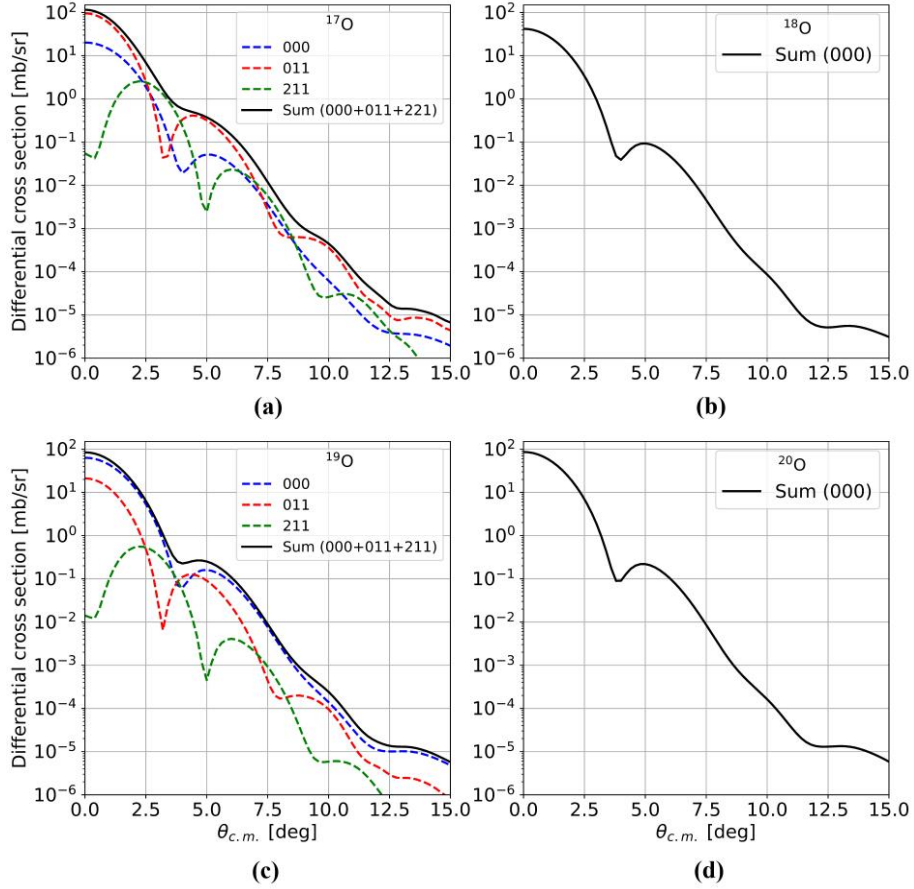

**Fig. 5** Differential cross sections for $^{17}$O($^3$He, $t$)$^{17}$F **(a)**, $^{18}$O($^3$He, $t$)$^{18}$F **(b)**, $^{19}$O($^3$He, $t$)$^{19}$F **(c)**, and $^{20}$O($^3$He, $t$)$^{20}$F **(d)** IAS reactions at 515 MeV/nucleon versus scattering angles in the center-of-mass frame. The blue dashed line represents the Fermi transition, the green dashed line represents the G-T transition, and the black solid line represents the sum in **(a)** and **(c)**. The black solid line represents the pure Fermi transition in **(b)** and **(d)**

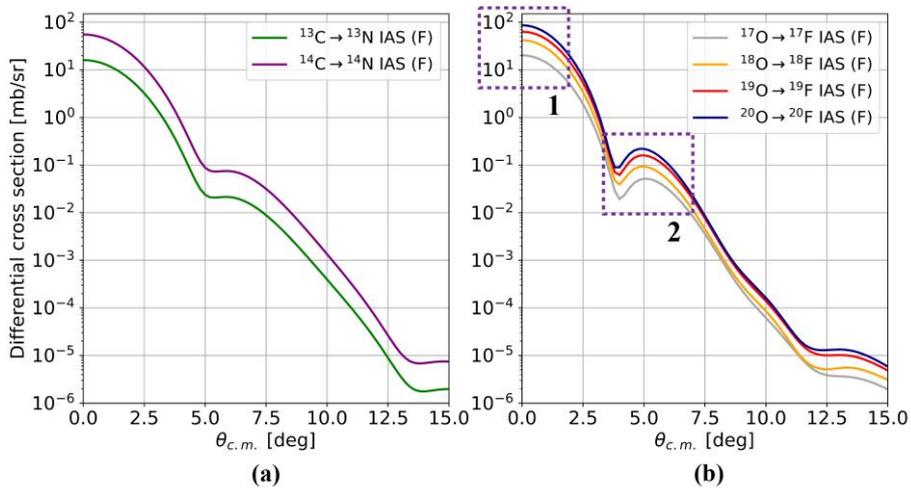

**Fig. 6** Differential cross sections for the CE ($^3$He, $t$) IAS reactions mediated by the Fermi transition of C and O isotopes at 515 MeV/nucleon versus scattering angles in the center-of-mass frame. **(a)** consists of $^{13,14}$C($^3$He, $t$) IAS reactions and **(b)** consists of $^{17,18,19,20}$O($^3$He, $t$) IAS reactions. In the purple box are two oscillating peaks

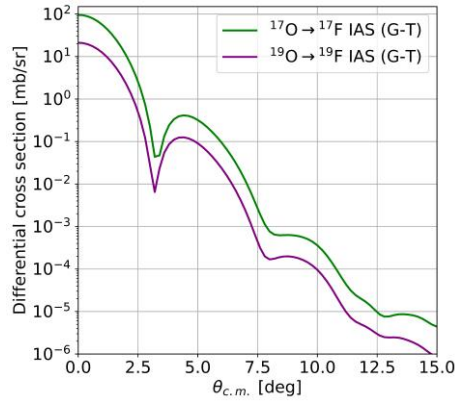

**Fig. 7** Differential cross sections for the CE ($^3$He, $t$) IAS reactions mediated by the G-T transition of $^{17}$O and $^{19}$O at 515 MeV/nucleon versus scattering angles in the center-of-mass frame

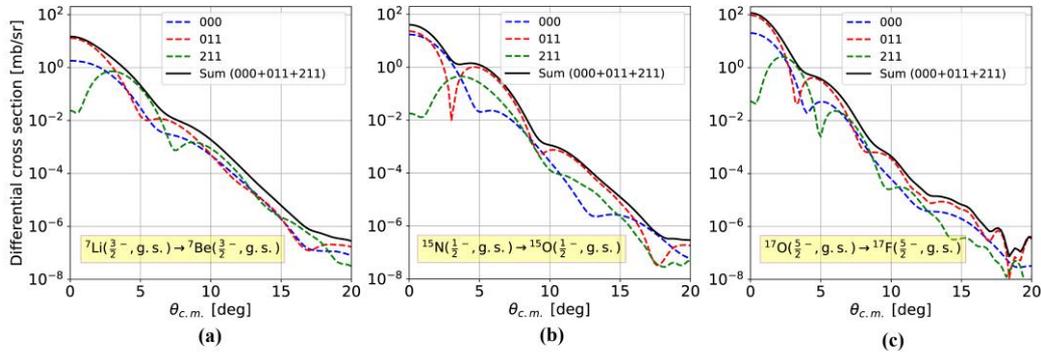

**Fig. 8** Differential cross sections for $^7$Li($^3$He, $t$)$^7$Be **(a)**, $^{15}$N($^3$He, $t$)$^{15}$O **(b)** and $^{17}$O($^3$He, $t$)$^{17}$F **(c)** IAS reactions at 515 MeV/nucleon versus scattering angles in the center-of-mass frame. The golden dashed line represents the Fermi transition, the green dashed line represents the G-T transition, and the solid blue line represents the sum

### 3.2 Identification of the IAS reaction

In ($^3$He, $t$) reaction experiments, the identification of IAS reactions primarily relies on the kinetic energy of the outgoing triton, which serves as a direct reflection of the energy level of the recoiled target nuclei. Besides, selecting a specific range of the scattering angle where the IAS reaction dominates is also valuable. In Fig. 9 and Fig. 10, the differential cross sections of major channels close to the IAS reaction for $^{13}$C($^3$He, $t$)$^{13}$N and $^{19}$O($^3$He, $t$)$^{19}$F are calculated as illustrations. In case of the excitation energy of the IAS reaction is significantly different from those of other reaction channels, such as the different 3 channels of $^{13}$C($^3$He, $t$)$^{13}$N shown in Fig. 9, it is convenient to distinguish the IAS reaction (0.00 MeV channel) by the kinetic energy of the triton. On the other side, the IAS reaction in $^{19}$O($^3$He, $t$)$^{19}$F (shown in Fig. 10) relates to an excited state of 7.904 MeV, which is difficult to be distinguished from the neighboring states (7.922 MeV and 7.446 MeV). However, different channels may correspond to different azimuthal distributions. By selecting an appropriate range of scattering angles, the dilution of other channels to the IAS reaction channel can be effectively eliminated. For the $^{19}$O($^3$He, $t$)$^{19}$F case, at the 0° scattering angle both the 7.922 MeV and 7.904 MeV channels are very strong and cannot be distinguished from each other. But as the scattering angle increases, the intensities of other channels rapidly decrease, especially for the 7.922 MeV channel and the IAS reaction gradually becomes the dominant component in a range between 5.0° to 10.0°, corresponding to the 2nd oscillation peak in Fig. 6**(b)**. In this region, although the other channels such as the one with an excited energy of 8.544 MeV shown in Fig. 10 also has a significant strength, by considering the energy difference of the outgoing triton the IAS reaction can be measured with precision. As the scattering angle increases further (at or above 10.0°), the IAS cross section is too small to be measured with a satisfied statistic uncertainty. In summary, to identify the IAS reaction over other channels, the kinematic region of the measurement should be carefully determined with a guide of the theoretical calculation mentioned above .

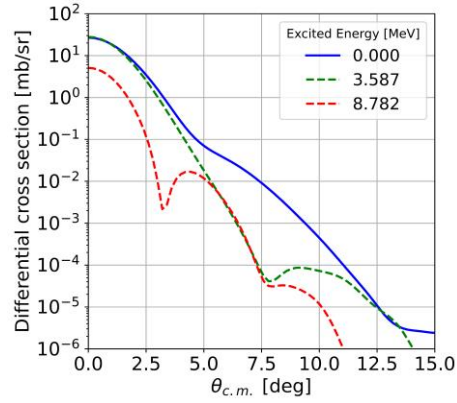

**Fig. 9** Differential cross sections versus scattering angles at different excitation energies of $^{13}$C($^3$He, $t$)$^{13}$N reaction, including several major reaction channels. For $^{13}$N, there are $J_f^\pi = 1/2^-$, $T_f = 1/2$ (0.000 MeV, 8.782 MeV) and $J_f^\pi = 3/2^-$, $T_f = 1/2$ (3.587 MeV)

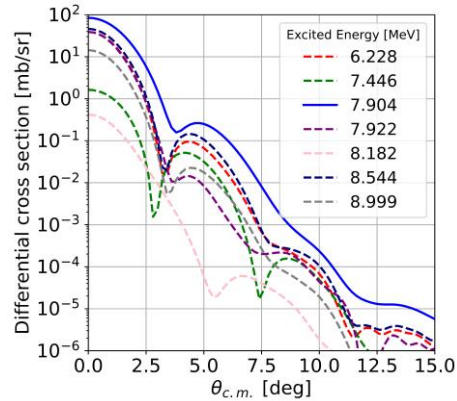

**Fig. 10** Differential cross sections versus scattering angles at different excitation energies of $^{19}$O($^3$He, $t$)$^{19}$F reaction, including several major reaction channels. For $^{19}$F, there are $J_f^\pi = 3/2^+$, $T_f = 1/2$ (7.446 MeV), $J_f^\pi = 3/2^+$, $T_f = 3/2$ (8.182 MeV), $J_f^\pi = 5/2^+$, $T_f = 1/2$ (6.228 MeV, 7.922 MeV, 8.544 MeV, 8.999 MeV) and $J_f^\pi = 5/2^+$, $T_f = 3/2$ (7.904 MeV)

3.3 Neutron radius

  The IAS reaction in CE reaction is closely related to the neutron-proton asymmetry of the nucleus, which can be represented as the difference between the proton radius ($R_p$) and neutron radius ($R_n$), $R_n$ - $R_p$. The proton radius $R_p$ has already been extensively measured by the scattering experiment and the atomic spectroscopy, and there are several empirical formulas precisely describing the charge distribution of the nuclei [61-66]. Meanwhile, the neutron distribution of the nucleus is extracted primarily by other different theoretical analyses [67] and various experimental programmes such as hadronic scattering [68], pion photo production [69], and parity-violating electron scattering [70]. It will be a strict global inspection of nuclear theory that the difference of the extracted proton radius and extracted neutron radius by different theoretical model is coincident with the observation of the IAS reaction in the CE reaction, especially on the nuclei far from the $\beta$-stable line. To demonstrate the connection between the difference of proton and neutron radii of a nuclei and the cross section of the IAS reaction, we employ a calculation with the `NuShellX` code. The `DENS` branch of `NuShellX` can export neutron and proton density distributions. `NuShellX` results are employed for neutron and proton radii to maintain a uniform standard. The $R_p$, $R_n$ and $R_n$ - $R_p$ are listed in Table 2. The differential IAS (via the Fermi transition) reaction cross sections versus $R_n$ - $R_p$ at 0° and 4.0° scattering angle in the center-of-mass frame are shown in Fig. 11.

**Table 2** Proton radii $R_p$, neutron radii $R_n$ and $R_n$ - $R_p$ calculated from the neutron and proton density distributions. The density distributions are exported by the `NuShellX`

|  | $R_p$ / fm | $R_n$ / fm | $R_n$ - $R_p$ / fm |
|---|---|---|---|
| $^{13}$C | 2.3138 | 2.3912 | 0.0774 |
| $^{14}$C | 2.3990 | 2.5293 | 0.1302 |
| $^{17}$O | 2.6253 | 2.7039 | 0.0786 |
| $^{18}$O | 2.6331 | 2.8119 | 0.1788 |
| $^{19}$O | 2.6415 | 2.9289 | 0.2874 |
| $^{20}$O | 2.6505 | 3.0339 | 0.3834 |

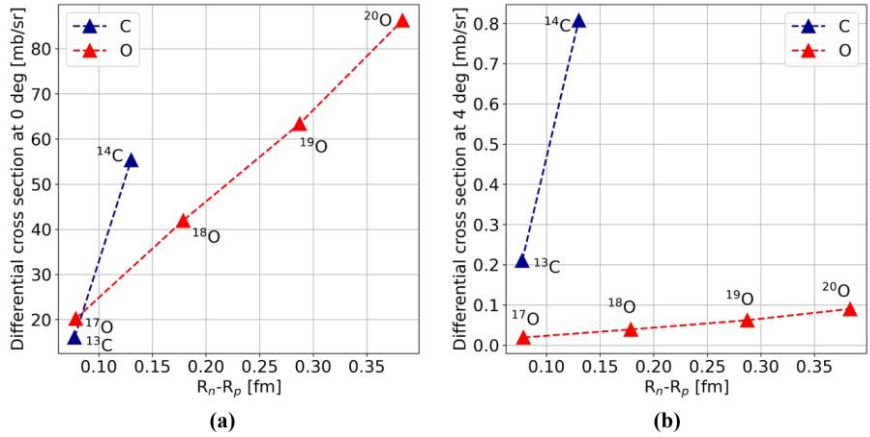

**Fig. 11** Differential IAS reactions cross sections (mediated by the Fermi transition) versus $R_n$ - $R_p$. The blue lines represent the C isotopes and the red lines represent the O isotopes. The differential cross sections at the 0° scattering angle in the center-of-mass frame are shown in **(a)**. The differential cross sections at the 4.0° scattering angle in the center-of-mass frame are shown in **(b)**

With the increasing neutron number along an isotope chain, there should be a distinct increase in the neutron radius, while only minor changes are expected in the proton radius. As shown in Fig. 6 and Fig. 11, with the increasing neutron radius, the IAS cross section mediated by the Fermi transition increases as well. This correlation can be clearly observed near oscillation peaks (marked as the dashed box at approximately 0° and 5.0° in Fig. 6**(b)**). Therefore, the measurement of IAS reactions near the oscillation peaks is expected to shed a light on the relationship between the IAS reaction and nuclear size.

3.4 Measurement in the inverse kinematics

In the experiment, inverse kinematics has been employed to observe the IAS reactions on the unstable isotopes using radioactive beams. In this work, the kinematic conditions of the experiment have been optimized to compromise the limited beam luminosity and energy, as well as the detector efficiency for each targeted isotope.

To choose a proper kinematic condition for the measurement, the relationship between the kinetic energies (i.e., the kinetic energy of the recoiled triton) and the scattering angle (in the center-of-mass frame) is first calculated (shown in Fig. 12). The reaction can be approximated as a two-body elastic scattering between a heavy ion and $^3$He. The relativistic kinematics is calculated, where the beam energy is in a range of 400 ~ 600 MeV/nucleon. Compared with the CE experiments in the normal kinematic region [1-3,26-32], the beam energy must be higher to observe the residual nuclei (the recoiled triton) (see Fig. 13). The triton needs sufficient kinetic energy to overcome the target's self-absorption. As shown in Fig. 12, within the scattering angle range of 5.0° to 10.0°, where the IAS reaction dominates, the kinetic energies of the triton are essentially the same for C and O beams with different energies. In the large-angle region (above 10.0°), the triton energies start varying with the ion type and beam energy. Therefore, the range of energies and scattering angles for different types of light nucleus and energies is essentially the same for the experiment. One detector design might fulfill the experimental requirements for the IAS reactions of light nucleus with an incident energy between 400 and 600 MeV/nucleon.

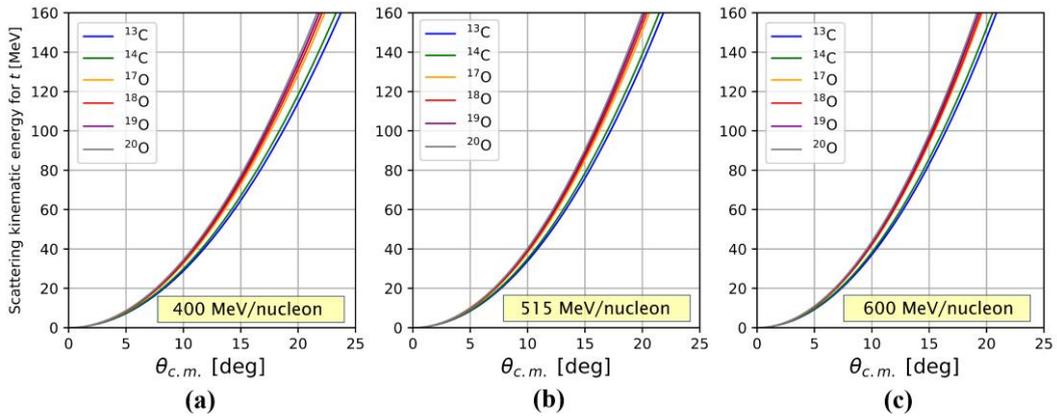

**Fig. 12** Scattering kinetic energy versus the scattering angle of the triton in the center-of-mass frame. The C and O beams with energies of 400 MeV/nucleon **(a)**, 515 MeV/nucleon **(b)**, and 600 MeV/nucleon **(c)** bombard the $^3$He target

A conceptional design of the proposed measurement is illustrated in Fig. 13. It is a $\Delta E$-$E$ telescope [71-72] composed by a Time Projection Chamber (TPC) and a scintillator array. The TPC, served as the $\Delta E$ detector, also offers a precise tracking detector. Since the CE reaction is in the quasi-elastic region, there is a strong correlation between the scattering angle and the kinetic energy of the outgoing triton. This correlation can be employed as a very handful selecting rule on the kinematics. On the other hand, the scintillator detector, which is composed of CsI(Tl) crystals and serves as the $E$ detector, also offers as an efficient trigger, which is necessary for the TPC. In this design, the range of scattering angle in the Laboratory frame is from 76.0° to 86.0°, corresponding to about 5.0° ~ 15.0° in the center-of-mass frame. The corresponding kinetic energy range is from 10 MeV to 140 MeV, which is limited by the thickness of the detector window and the scintillator size. The $\Delta E$-$E$ measurement offers a clean particle identification over all of the outgoing particles in the same scattering angle. In order to demonstrate the triton identification, a primary simulation based on Geant4 [73-75] toolkit was done. In the simulation, the $^3$He target was bombarded with a 515 MeV/nucleon $^{17}$C beam. The $\Delta E$-$E$ distribution is shown in Fig. 14. The collision process is based on the FTFP_BERT_ATL physics list [76], which only give a statistical description about the scattering process.

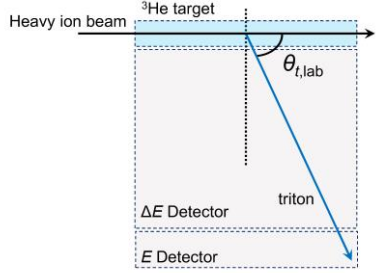

**Fig. 13** A conceptional design of the measurement for the CE reaction

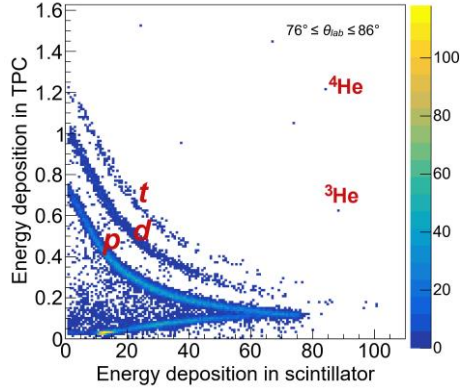

**Fig. 14** Simulated $\Delta E$-$E$ distribution of products from the bombardment of the $^3$He target by the 500 MeV/nucleon $^{17}$C beam

Taking the detector acceptance into consideration, the macroscopic cross section $\sigma_{\text{IAS}}$ of the CE reaction can be expressed as:

$$\sigma_{\text{IAS}} = \int_{\varphi_1}^{\varphi_2} d\varphi \int_{\theta_1}^{\theta_2} \sin\theta \frac{d\sigma}{d\Omega}(\theta)|_{\text{IAS}} d\theta, \tag{7}$$

where $\varphi$ is the azimuthal angle, $\theta$ is the scattering angle, and $d\sigma/d\Omega$ is the differential cross section. The rate of yield $N_{\text{IAS}}$ can be estimated as:

$$N_{\text{IAS}} = \sigma_{\text{IAS}} I N_s, \tag{8}$$

where $I$ is the beam intensity and $N_s$ is the number density of the $^3$He atoms in the target. Assuming the $\varphi$ angle ranges from 0° to 180.0°, the $\theta$ angle ranges from 76.0° to 86.0°, the beam intensity is $10^6$ particles per second, and the target density is 2 amagats, the yield rates of the triton in IAS reactions are calculated by Eq. **(8)**, shown as Table 3. According to the estimated rates, it is possible to get enough statistics within a limited beam time.

Table 3 Prediction cross sections and counting rates of CE IAS reactions

| | Cross section / μb | Counting rate / h$^{-1}$ |
|---|---|---|
| $^{13}$C($^3$He, $t$)$^{13}$N (g.s.) | 10.79 | 55.70 |
| $^{14}$C($^3$He, $t$)$^{14}$N (2.690 MeV) | 11.04 | 57.02 |
| $^{17}$O($^3$He, $t$)$^{17}$F (g.s.) | 4.54 | 23.44 |
| $^{18}$O($^3$He, $t$)$^{18}$F (1.592 MeV) | 1.76 | 9.10 |
| $^{19}$O($^3$He, $t$)$^{19}$F (7.904 MeV) | 3.88 | 20.08 |
| $^{20}$O($^3$He, $t$)$^{20}$F (6.979 MeV) | 3.68 | 19.02 |

## 4  Conclusion

In this work the differential cross sections for charge exchange reactions are calculated in the framework of the double-folding potential and distorted-wave Born approximation. In this theorical framework the $^{13,14}$C($^3$He, $t$) and $^{17,18,19,20}$O($^3$He, $t$) reactions are investigated, specifically focusing on the relationship between the outer shell structure and the IAS reactions channels.

The IAS reactions for different nucleus may consist of various components. The $^{13}$C($^3$He, $t$) and $^{17,19}$O($^3$He, $t$) IAS reactions involve both the Fermi and the G-T transitions, while the $^{14}$C($^3$He, $t$) and $^{18,20}$O($^3$He, $t$) IAS reactions are mediated by the Fermi transition only. In addition to the established dependence of the Fermi transition in the IAS reaction on the neutron radius of the even-even isotopes, a similar correlation between the G-T transition and the neutron radius was also observed for the odd-A isotopes such as $^{13}$C and $^{17,19}$O. More interestingly, in the odd-A cases, it was observed that the G-T transition depends not only on the excess neutron but also on the angular momentum configuration of the outer neutrons. This feature might be utilized to explore the nuclear clustering phenomenon of the light neutron-rich isotopes, such as $^7$Li, $^{15}$N, and $^{17}$O. Therefore, conducting accurate ($^3$He, $t$) experiment to extract the IAS reactions of unstable nucleus will provide new insights into their isospin structure. For such a measurement, a range of scattering angle from 5.0° to 10.0° in the center-of-mass frame would be ideal, where the IAS reactions dominate and can be conveniently distinguished from other reaction channels. For nuclei such as $^{13}$C and $^{17,19}$O, the extraction of both Fermi and G-T transition from the total IAS reaction cross section by the MDA was demonstrated, according to the calculated reaction cross section.

In this work, the inverse kinematics of the CE reaction was also discussed, with the beam condition of the HIRFL-CSR. For incident light nuclei with the intermediate energy, the operating range of the detector was determined to be within 76.0°~86.0° in the laboratory frame, corresponding to a range from 5.0° to 15.0° in the center-of-mass frame. The yields of the triton and the byproducts (proton and deuteron) are estimated. And it is demonstrated that enough statistics can be achieved within a limited beam time and luminosity.

In the future, the experiment will be carried out. And more nuclei and isotopic chains will be calculated to investigate the nuclear structure involved in the G-T transition more carefully.

**Acknowledge** This work is financially supported by National Key R&D Program of China (Grant No. 2022YFE0103900) and the National Natural Science Foundation of China (Grant Nos., U2032166, 11875301, and U1832167).